\documentclass[runningheads]{llncs}

\usepackage[T1]{fontenc}
\usepackage{graphicx}
\usepackage{booktabs}
\usepackage{amsmath,amssymb,amsfonts}
\usepackage[misc]{ifsym}
\usepackage{hyperref}
\usepackage{url}            
\usepackage{subcaption}
\usepackage{multirow}
\usepackage{float}
\usepackage[normalem]{ulem}
\usepackage{enumitem}
\usepackage{comment}

\usepackage{pifont}
\newcommand{\cmark}{\ding{51}} 
\newcommand{\xmark}{\ding{55}} 

\hypersetup{
   colorlinks,
   menucolor=black,
   linkcolor=black,
   citecolor=red,
   urlcolor=blue
}

\begin{document}

\title{A Critical Review on the Effectiveness and Privacy Threats of Membership Inference Attacks}

\titlerunning{A Critical Review on the Effectiveness and Privacy Threats of MIAs}

\author{Najeeb Jebreel
\and David S\'anchez 
\and Josep Domingo-Ferrer}

\authorrunning{Najeeb Jebreel {\em et al.}}

\institute{Universitat Rovira i Virgili,\\ Department of Computer Engineering and Mathematics,\\
        CYBERCAT-Center for Cybersecurity Research of Catalonia,\\ ComSCIAM-Center for Computational Science and Applied Mathematics \\ Av. Pa\"{\i}sos Catalans 26, 43007 Tarragona, Catalonia\\
\email{\{najeeb.jebreel, david.sanchez, josep.domingo\}@urv.cat}}
\maketitle

\begin{abstract}
Membership inference attacks (MIAs) aim to determine whether a data sample was included in a machine learning (ML) model's training set and have become the {\em de facto} standard for measuring privacy leakages in ML. 
We propose an evaluation framework that defines the conditions under which MIAs constitute a genuine privacy threat, and review representative MIAs against it. We find that, under the realistic conditions defined in our framework, MIAs represent weak privacy threats. 
Thus, relying on them as a privacy metric in ML can lead to an overestimation of risk and to unnecessary sacrifices in model utility as a consequence of employing too strong defenses.

\end{abstract}

\section{Introduction}
\label{sec:introduction}
Advancements in machine learning (ML) have significantly improved performance in many tasks~\cite{devlin2019bert,miotto2018deep}.
However, these advances have also raised  privacy concerns among individuals and regulatory authorities due to the use of potentially sensitive data to train ML models.

The most widespread concern is the vulnerability of ML models to membership inference attacks (MIAs)~\cite{shokri2017membership,yeom2018privacy,salem2019ml}. 
MIAs aim to determine whether a specific data point was included in a model's training data set by exploiting differences in model behavior (such as loss or confidence scores) when presented with member and non-member data points.
Indeed, in some scenarios, successfully inferring an individual's membership in a training data set can compromise privacy by revealing sensitive information.
For example, if an ML model is trained on medical records of patients with a common sensitive condition, such as HIV or cancer, confirming membership would expose the individual's sensitive health status.
Recognizing this risk, organizations such as NIST (USA) and ICO (UK)~\cite{tabassi2019taxonomy,murakonda2020ml} as well as researchers 
in the health domain~\cite{pilgram2025privacy} consider MIAs a possible violation of training data confidentiality.

Driven by these concerns, and to comply with privacy protection laws such as the European Union's General Data Protection Regulation (GDPR)~\cite{gdpr}, several defenses against MIAs have been proposed. Differential privacy (DP)~\cite{dwork2006differential} is the most widespread and natural defense, as it obscures the influence of any 
individual data point in the training data on the trained ML model by adding random noise. 
However, DP reduces the utility of the model due to the noise required to achieve meaningful privacy protection~\cite{bagdasaryan2019differential,runhua2021ppml,blanco2022critical}.

On the other hand, early work on MIAs by \cite{yeom2018privacy} established a connection between the vulnerability of a model to MIAs and the degree to which the model overfits the training data. 
Indeed, it seems reasonable that a model that is overfitted on its training data will reveal more about them than a model that has learned more general patterns.
However, overfitting is undesirable in production ML because
it reduces the ability of models to generalize.
Moreover, as has been acknowledged in the privacy literature for decades \cite{sdc-book}, for a disclosure inference ---such as membership--- to involve a real privacy threat, that inference should be unequivocal, which is something very rare in MIAs. 

A critical question arises: \emph{do membership inference attacks pose a significant threat to privacy under realistic conditions, thereby justifying the adoption of defenses such as DP despite their utility downsides?}

\paragraph{\bf Related work.} 
Several studies have critically examined the effectiveness of MIAs. 
The literature has established a strong connection between model overfitting and vulnerability to MIAs~\cite{shokri2017membership,yeom2018privacy,salem2019ml}.
Although MIAs can be effective against overfitted models, their performance deteriorates when applied to well-generalized models~\cite{dionysiou2023sok}. 
However, even with non-overfitted models, some samples may remain vulnerable to MIAs~\cite{long2018understanding,carlini2022membership,zarifzadeh2024low}.
It is shown in \cite{truex2021demystifying} that MIAs work well under specific conditions, such as when the data set is complex, the model is sensitive to individual instances, and the attacker can generate accurate approximations of the target model. In
\cite{duan2024membership,he2024difficulty} it is demonstrated that MIAs tend to barely outperform random guessing for large-scale models trained on big and comprehensive data. 
In \cite{jebreel2026revisiting}, it is shown that the effectiveness of the strong LiRA MIA~\cite{carlini2022membership}
collapses under a realistic evaluation that combines anti-overfitting 
training, shadow-based threshold calibration, and skewed membership priors, 
with poor reproducibility of inferred vulnerable samples across 
runs.
In~\cite{aubinais2025fundamental} it is theoretically shown that the effectiveness of MIAs is inherently constrained by the statistical properties of the training data. It is argued in 
\cite{rezaei2021difficulty,rezaei2022discredibility} that MIAs are unreliable because they exhibit high false positives due to the frequent misclassification of semantically similar non-members as members. 
Adjusting the precision to account for realistic membership
priors is proposed in~\cite{jayaraman2021revisiting}. 
Recently, a consensus has emerged to use the true positive
rate (TPR) at an extremely low false positive rate (FPR) as a benchmark to ensure MIA reliability~\cite{carlini2022membership,zarifzadeh2024low,he2024difficulty}.

\paragraph{\bf Our contributions.} Although the related work mentioned above provides valuable insight into the effectiveness of MIAs, there is a gap to understand the conditions under which MIAs constitute a meaningful privacy threat and how realistic these conditions are. In this work, we address this gap with the following contributions:
\begin{itemize}
    \item We propose an evaluation framework that defines the necessary conditions for MIAs to be considered genuine privacy threats in predictive ML ---the scenario for which MIAs were designed and on which the surveyed works focus. This framework assesses MIAs along five dimensions:
    disclosure potential of the target model,
    applicability to non-overfitted models, applicability to models that are competitive for real-world deployment, attack reliability, and computational feasibility.
    \item We review representative MIAs through the lens of this framework, evaluating their effectiveness and implications for real-world privacy.
    \item We discuss the role of MIAs in the assessment of privacy risk and whether their impact justifies the adoption of utility-hampering defenses such as DP.
\end{itemize}
Our findings reveal that, despite their apparent success in controlled settings, 
existing MIAs on realistic training data sets and ML models do not meet the conditions required for meaningful privacy threats in real-world scenarios. 
Thus, relying on MIAs as the primary metric for privacy risk in predictive ML may lead to overestimated threats and unnecessary sacrifices in model utility.

The remainder of this paper is organized as follows. 
Section~\ref{sec:disclosure_risk} examines MIAs from the point of view of disclosure risk.
Section~\ref{sec:framework} presents our framework for evaluating when MIAs pose a genuine privacy threat. 
Section~\ref{sec:evaluation} reviews representative MIAs against our evaluation framework. 
Section~\ref{sec:discussion} discusses the wider implications of our findings for privacy and ML utility. 
Finally, Section~\ref{sec:conclusion} provides concluding remarks.

\section{MIAs and Disclosure Risk}
\label{sec:disclosure_risk}

The concept of \emph{disclosure risk} has long served as a foundation for understanding how sensitive information can be accidentally exposed in data sets released. 
Originally developed in the context of database privacy \cite{sdc-book}, its principles remain relevant in the machine learning domain. 

There are two types of disclosure~\cite{sdc-book}:
i) \emph{identity disclosure (a.k.a. re-ident\-ifica\-tion)}, which allows associating a released non-identified record with the subject to whom it corresponds; and
ii) \emph{attribute disclosure}, which enables determining the value of a subject’s confidential attribute, such as income or diagnosis.

Attributes that enable direct re-identification are \emph{personal identifiers} (such as passport numbers), whereas \emph{quasi-identifiers} (\emph{e.g.}, zip code, gender or age) are those that do not uniquely identify the subject, but whose combination may because they may be present in public identified databases like electoral rolls. Finally, \emph{confidential} attributes are unknown information on subjects that might reveal their sensitive data (\emph{e.g.}, salaries, diagnoses) when unequivocally associated with them. 

Unlike re-identification attacks, MIAs are not designed to identify the subject's record directly. 
Instead, they focus on determining whether a given data sample was included in an ML model training set. However, disclosure of training membership can indirectly disclose sensitive information in two ways:
\begin{enumerate}
    \item \emph{Revealing confidential attributes:} If a data sample explicitly contains confidential features or labels, confirming its presence in the training set may reveal private information about the individual. For example, if a data set contains medical records, confirming that a person's data is included may reveal a sensitive diagnosis.
    \item \emph{Exposing confidential contexts:}  Membership in a training data set associated with a sensitive context (for example, mental health studies, addiction treatment programs) can itself constitute a violation of privacy, even if there are no explicit sensitive attributes present. In fact, this scenario can be viewed as a training data set with a single homogeneous confidential value that is shared by all members. 
\end{enumerate}
To pose a privacy threat, MIAs must accurately and unequivocally disclose \emph{new} information beyond what is already known from population-level statistics; or, in other words, 
for an MIA to be effective against privacy, it should significantly shift the attacker's prior belief about the attributes of a data point or its association with a confidential context to a different posterior belief.

However, there is a clash between membership disclosure and 
attribute disclosure, for the following reasons:
\begin{itemize}
    \item A necessary condition for unequivocal attribute disclosure is that training data must be an exhaustive representation of a population. Otherwise,  
    the attacker cannot be sure that the targeted subject was truly a member of the training data, as their known information could be shared by several other individuals in the population (present or not in the training data). That is, in 
    non-exhaustive population samples, there is \emph{plausible deniability} of any disclosure inference.
    \item Conversely, if training data are an exhaustive representation of the population (for example, a country-level census), then membership disclosure is trivial: everyone is known to be a member.
\end{itemize}

In fact, {\em exhaustivity} is not the only necessary condition for unequivocal attribute inference through MIAs. {\em Uniqueness of confidential attribute values} is also needed, which means that there should {\em not} be two or more records in the data set that: i) match the target subject's attributes known to the attacker; ii) have different values for the confidential attribute
values the attacker wishes to infer. Without uniqueness, no unequivocal inference of those attributes is possible
to the attacker. 
Finally, another necessary condition is that the information assumed to be known to the attacker about the target should be plausible. Assuming knowledge of too specific or too much information on the subject makes disclosure attacks contrived and unrealistic. 


In summary, for an MIA to be considered a genuine privacy threat, it must not only infer membership with very high or, ideally, perfect precision, but the underlying data must also fulfill strict (and often improbable) conditions to allow unequivocal disclosure of sensitive information. In particular, \textit{unequivocal attribute disclosure occurs only when membership disclosure is trivial because the data set is exhaustive}.

\section{Proposed Evaluation Framework}
\label{sec:framework}
\begin{figure}[t]
        \centering
        \includegraphics[width=0.7\textwidth]{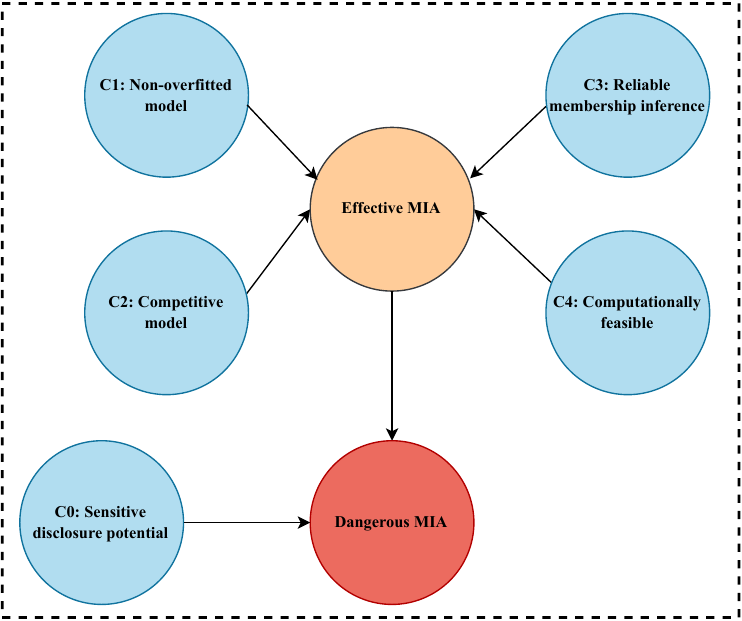}
        \caption{Overview of the proposed evaluation framework for MIAs. Condition C0 relates to the MIA disclosure potential, which critically depends on the data set used to train the target model. Conditions C1--C4 characterize the effectiveness of the MIA itself, which should reliably attack non-overfitted and competitive models at a reasonable computational cost.    
        }
        \label{fig:framework}
    \end{figure}

Building on the discussion so far, we propose an evaluation framework that defines five necessary conditions (C0–C4) that must hold simultaneously for an MIA to be a real threat to privacy in  predictive ML. 
C0 is a data-level precondition independent of the attack design, while C1–C4 assess the attack itself.
Figure~\ref{fig:framework} provides a schematic overview of the conditions and their implications.

\paragraph{\textbf{Condition C0: Sensitive disclosure potential.}}
\label{sec:condition5}
This condition requires the attack to have the potential to bring about a meaningful disclosure of sensitive data.
According to the discussion in Section~\ref{sec:disclosure_risk}, this depends on the data set on which the target model has been trained and requires i) {\em the training data to be an exhaustive sample of a population}, ii) {\em uniqueness of confidential attribute values}, and iii) {\em plausibility on information assumed to be known to the attacker about the target}.
C0 can be viewed as a {\em precondition that is agnostic of the precise design of the MIA attack}. If C0 does not hold, then an MIA cannot succeed, no matter how well it is crafted. 

\paragraph{\textbf{Condition C1: Non-overfitted model.}}
\label{sec:condition1}
An ML model is said to overfit when its average performance ({\em e.g.}, loss or accuracy) on the training data significantly exceeds its performance on unseen data from the same population, \emph{i.e.}, there is a large generalization gap~\cite{shalev2014understanding}. 
Overfitting occurs primarily when the model not only learns general patterns but also memorizes sample-specific details and noise, resulting in distinct behaviors for training data (\emph{members}) versus unseen data (\emph{non-members}).
In such cases, MIAs can trivially distinguish between members and non-members~\cite{yeom2018privacy,salem2019ml}.
However, for an MIA to be considered effective, it must succeed against non-overfitted models~\cite{carlini2022membership,zarifzadeh2024low}, which are the desirable ones in production settings. 

To avoid any margin of doubt, we adopt quite a permissive threshold that considers a model as ``non-overfitted'' whenever the train-test accuracy gap (\emph{i.e.}, $g = \mathrm{Acc}_{\text{train}} - \mathrm{Acc}_{\text{test}}$) is $g \le 10\%$.
Notice that $g > 10$ clearly indicates substantial overfitting or poor generalization.

\paragraph{\textbf{Condition C2: Competitive 
model.}}
\label{sec:condition2}
For an MIA to be considered meaningful, it must target a model that could realistically be deployed in real-world applications and thus be accessible to potential attackers. 
This necessitates that the model's utility be competitive with state-of-the-art (SotA) benchmarks for its main task. 
Models that fail to meet this criterion are unlikely to be used in production and hence do not reflect practical privacy risks.

We consider a model to be \emph{competitive} if its test accuracy is not 
more than $5\%$ lower than that of its SotA ---which is not much to ask---, \emph{i.e.},
$\Delta = \mathrm{Acc}_{\text{SotA}} - \mathrm{Acc}_{\text{test}} \le 0.05 \times \mathrm{Acc}_{\text{SotA}}$.
Models falling outside this range are unlikely to be selected for deployment when higher-performing alternatives are available.

Although the specific thresholds chosen for C1 and C2 can be debatable, our choices are deliberately permissive ({\em e.g.}, allowing up to 10\% generalization gap for C1, or 5\% deviation from SotA for C2). 
Because no one can reasonably argue that a model violating these thresholds is non-overfitted or competitive, stricter thresholds would only strengthen our conclusions.

\paragraph{\textbf{Condition C3: Reliable membership inference.}}
\label{sec:condition3}
An effective MIA must reliably distinguish members from non-members under realistic conditions.
Recent state-of-the-art works~\cite{ye2022enhanced,carlini2022membership,bertran2023scalable,zarifzadeh2024low}
focus on measuring the true positive rate (TPR) at an extremely low false positive rate (FPR) (FPR $\leq 0.1\%$) to ensure the
reliability of positive inferences.
However, this approach overlooks the typically low prior probability of
membership, since the training set often represents only a small subset of the
overall population.
As noted in~\cite{jayaraman2021revisiting}, standard precision does not account
for this imbalance.

To address this issue, a weighted precision metric that incorporates the prior
membership probability $p$ was proposed in~\cite{jayaraman2021revisiting}:
\begin{equation}
\label{eq:prec}
\text{Prec} = \frac{p \times \text{TPR}}{p \times \text{TPR} + (1-p) \times \text{FPR}},
\end{equation}
where $p \ll 50\%$ in realistic settings.

We assess reliability using two criteria:
\begin{enumerate}
    \item The attack must identify some true members with an FPR near 0\%. 
    This is crucial because non-members largely outnumber members in practice, and even small FPR values can lead to many false positives.
    \item The weighted precision \(\text{Prec}\) (as defined above in Equation~\eqref{eq:prec}) must be near perfect, indicating that the positive inferences are indeed true members, even when accounting for low realistic membership priors. 
    Weighted precision also relates to plausible deniability~\cite{bindschaedler2017plausible}, a privacy concept that ensures that an attacker cannot definitively determine whether an individual's data are included in a data set. In the context of MIAs, plausible deniability can be claimed if the attack is imperfect. The level of plausible deniability is inversely proportional to the attack's precision. 
    We require a precision $\ge 95\%$ to counter plausible deniability claims. 
    This ensures that at most 5\% of flagged samples are false positives. Although this does not directly correspond to statistical significance ($\alpha = 5\%$), it aligns with a standard threshold to reduce erroneous identifications.
\end{enumerate}

We evaluate the precision with a realistic membership prior $p=10\%$, which is
already conservative relative to many deployment scenarios.
For some data sets (\emph{e.g.},~\cite{jayaraman2021revisiting}), even lower priors are
considered.

\paragraph{\textbf{Condition C4: Computational feasibility.}}
\label{sec:condition4}

Even if an attack is reliable against non-overfitted competitive models, 
it must be feasible with the computational resources reasonably available to potential attackers. 
This is in line with legal frameworks such as the GDPR (Recital 26)~\cite{gdpr}, which states that the evaluation of disclosure risk must account for the means reasonably likely to be used in an attack, including cost, time, and available technology. 

An attack whose computational cost matches or exceeds that of training the target model is disproportionate, as the attacker’s effort would rival or surpass the defender’s. 
In such cases, the attack becomes largely unaffordable in realistic settings. 
We quantify the computational demands of MIAs through three resource factors, ordered by their contribution to overall cost.

The first and most significant factor is the number of additional models ($M$) ---\emph{e.g.}, shadow, distilled, or reference models--- required by the attack: training those models represents a major computational overhead, which can even exceed the effort devoted to building the original target model. 
The cost is considered \emph{low} when no additional models are required ($M = 0$), \emph{moderate} when a single additional model is needed ($M = 1$) --with complexity on par with the target model--, and \emph{high} when multiple additional models are required ($M \geq 2$).

The second factor is the cost of the inference model ($I$). 
This cost is categorized as \emph{low} when a computationally trivial rule is used (\emph{e.g.}, a simple threshold on loss values), \emph{moderate} when a moderately complex model is employed (\emph{e.g.}, a simple binary classifier), and \emph{high} when a computationally expensive inference mechanism is required (\emph{e.g.}, deep neural networks such as~\cite{bertran2023scalable}).

The third factor is the number of model queries required per target sample ($Q$). 
Although not as dominant as $M$ and $I$, the query count per sample still contributes to the overall cost. 
This requirement is considered \emph{low} when a maximum of 100 queries per target sample is required, \emph{moderate} when between 101--500 queries are needed, and \emph{high} when a larger number is required. Notice that larger numbers also increase the risk of attack detection.

\textbf{Aggregation into overall cost levels.} 
The three independently assessed factors are finally aggregated into an overall cost level. 
A \emph{low-cost} MIA does not require additional models ($M = 0$), uses a low-complexity inference rule, and requires a low number of queries per sample. 
An example is the simple threshold-based attack by \cite{yeom2018privacy} performed on a target model with a single query per sample.

A \emph{moderate-cost} MIA requires a single additional model ($M = 1$), and/or an inference model of moderate complexity and/or a low-moderate number of queries per sample. 
An example is an attack that trains one shadow model and employs a shallow classifier for inference with 10-200 queries per target sample.

A \emph{high-cost} MIA requires multiple additional models ($M \geq 2$), and/or a high-complexity inference model regardless of the query cost, and/or a very high number of queries per sample when coupled with even moderate levels of the other factors.

\section{Review and Evaluation of Representative MIAs}
\label{sec:evaluation}

This section evaluates, under the lens of our framework, key contributions on black-box MIAs against deep neural networks (DNNs) used for classification. The latter is by far the most common scenario considered by the surveyed works.
We focus on the most influential and representative attacks in the literature, specifically, those presented in top-tier venues and that have attracted the highest number of citations (13,738 citations overall as of January~8,~2026). 
To gain a broader perspective and analyze their evolution, we include both pioneering (such as \cite{shokri2017membership}) and recent state-of-the-art attacks (such as LiRA~\cite{carlini2022membership}, \cite{bertran2023scalable}, and \cite{zarifzadeh2024low}).
Detailed descriptions of the surveyed attacks are provided in Appendix~\ref{supp:surveyed-attacks}.

\paragraph{\textbf{Condition C0.}}
In Section~\ref{sec:disclosure_risk}, we highlighted that for non-trivial membership disclosure to be possible, the training data set should not be an exhaustive sample of the population. 

We assess condition C0 by analyzing the data sets used for the classification tasks considered by the MIAs surveyed.
{\em Tabular data sets} include Adult~\cite{adult_2}, 
Purchase-100~\cite{shokri2017membership},
Texas-100~\cite{shokri2017membership},
UCI Credit~\cite{credit_data},
UCI Hepatitis~\cite{hepatitis_46}, and
UCI Cancer~\cite{breast_cancer}.
{\em Image data sets} are 
widely used to evaluate MIAs and they range from simple digit recognition (MNIST) to complex object detection 
(ImageNet). These are:  
MNIST~\cite{lecun1998gradient}, CIFAR-10~\cite{krizhevsky2009learning}, CIFAR-100~\cite{krizhevsky2009learning}, CINIC-10~\cite{darlow2018cinic}, GTSRB (German Traffic Sign Recognition Benchmark)~\cite{stallkamp2011german}, ImageNet-1K~\cite{russakovsky2015imagenet} (a sample from ImageNet~\cite{deng2009imagenet}), and LFW (Labeled Faces in the Wild)~\cite{huang2007lfw}.
{\em Textual data sets} include Newsgroups~\footnote{\url{https://scikit-learn.org/0.19/data sets/twenty_newsgroups.html}} and 
RCV1X (the Ms RCV1 data set)~\cite{lewis2004rcv1}.
Finally, we consider the Locations \emph{trajectory data set}~\cite{shokri2017membership}.

None of these data sets is an exhaustive sample of the corresponding population, namely, American citizens for Adult, shoppers for Purchase-100, patients for Texas-100, bank customers for UCI Credit, hepatitis patients for UCI Hepatitis, cancerous cells for UCI Cancer, locations of Foursquare users for Locations, images of handwritten digits for MNIST, images in general for CIFAR-10, CIFAR-100, CINIC-10, and ImageNet-1K, German traffic sign images for GTSRB, face images for LFW, text documents for Newsgroups and RCV1X. 
Hence, {\em non-trivial membership disclosure is {\em a priori} possible.}

However, with respect to potential attribute disclosure:
\begin{itemize}
    \item We explain in Section~\ref{sec:disclosure_risk} that
unequivocal attribute disclosure is not possible when training data are not exhaustive. Therefore, {\em none of the reviewed MIAs can result in unequivocal attribute disclosure.}
\item Most of the data sets contain public non-sensitive data (MNIST, CIFAR-10, CIFAR-100, ImageNet-1k, CINIC-10, GTSRB, RCV1X, and Newsgroups), which is understandable for reproducibility and to enable comparative analyses. However, even if we consider that some of their attributes might be confidential, membership would only allow inferring them if those data sets were exhaustive and satisfied uniqueness of the allegedly confidential attribute values. That is, 
the MIA attacker should check that there are no two or more records in the data set that match the attribute values known to the attacker but have different values for the unknown (confidential) attributes. None of the attacks reviewed documents or even considers attribute uniqueness. 
\end{itemize}

Regarding C0, we can conclude that i) the training data sets used are not exhaustive,
which means that MIAs are non-trivial, but their results are not unequivocal, and ii) 
the lack of uniqueness of confidential attributes is a relatively outlying condition that should not significantly affect most data sets used.

\paragraph{\textbf{Condition C1.}}
The first columns of Table~\ref{tab:model_realism} report the evaluation of the model overfit on all the pairs of attack-data sets surveyed. Among the 61 pairs, 24 ($39.34\%$) lack sufficient information
to assess overfitting (NA).
Among the remaining 37 pairs, 27 ($72.97\%$) use overfitted target models
with train-test gaps exceeding $10\%$, while only 10 ($27.03\%$) satisfy C1.

This behavior is strongly data set-dependent.
Simple benchmarks such as Adult and MNIST often meet C1, while complex vision
benchmarks (CIFAR-10/100, ImageNet-1K) are frequently evaluated using models with
large generalization gaps.
Importantly, most violations of C1 are not marginal: many gaps substantially
exceed $10\%$, indicating pronounced overfitting rather than borderline cases.

\paragraph{\textbf{Condition C2.}}
The evaluation of model competitiveness is reported in the last columns of Table~\ref{tab:model_realism}.
The required test accuracy is missing for 11 pairs ($18.03\%$).
Among the remaining 50 pairs, only 12 ($24.00\%$) use target models within
$5\%$ of the SotA.
The vast majority (38 pairs, $76.00\%$) evaluate MIAs on models that lag far
behind SotA, often by large margins, particularly on CIFAR-10/100 and
ImageNet-1K.

Taking into account only the 37 pairs for which C1 was assessable, just 7 pairs
($18.92\%$) satisfy both C1 and C2 simultaneously.
Most evaluations fail to meet at least one of the two conditions, due to model overfitting or non-competitive target models.
As a result, the MIA effectiveness reported often reflects unduly favorable experimental
conditions rather than realistic deployment scenarios.

All in all, the results in Table~\ref{tab:model_realism} indicate that a large
fraction of the MIA literature evaluates attacks on target models that are not
representative of well generalized and competitive systems. 
This inflates the
apparent success of attacks and weakens their relevance for practical
privacy risk assessment.

\begin{table}[t!]
\centering
\caption{Evaluation of model non-overfitting (C1) and competitiveness (C2)}
\label{tab:model_realism}
\resizebox{0.84\textwidth}{!}{%
\begin{tabular}{ll|cccc|cccc}
\toprule
Attack & Data set &
Train acc. (\%) & Test acc. (\%) & $g$ (\%) & Non-overfitted? &
SotA acc.  (\%) & $\Delta$ (\%) & Competitive? \\
\midrule

\multirow{7}{*}{\cite{shokri2017membership}}
& Adult        & 84.80 & 84.20 & 0.60  & \cmark & 88.16~\cite{chakrabarty2018statistical} & 3.96  & \cmark \\
& Purchase-100 & NA    & 72.90 & NA    & NA     & 90.00~\cite{suri2024parameters} & 17.10 & \xmark \\
& Texas-100    & NA    & 57.00 & NA    & NA     & 57.00~\cite{shokri2017membership} & 0.00  & \cmark \\
& Locations    & 100.00& 67.30 & 32.70 & \xmark & 72.00 \cite{li2024mist} & 4.70  & \cmark \\
& MNIST        & 98.40 & 92.80 & 5.60  & \cmark & 99.87 \cite{byerly2021no} & 7.07  & \xmark \\
& CIFAR-10     & NA    & 60.00 & NA    & NA     & 99.50 \cite{dosovitskiy2021an} & 39.50 & \xmark \\
& CIFAR-100    & NA    & 20.00 & NA    & NA     & 96.08 \cite{foret2021sharpnessaware} & 76.08 & \xmark \\
\midrule

\multirow{3}{*}{\cite{yeom2018privacy}}
& MNIST        & NA & NA & NA & NA & 99.87 \cite{byerly2021no} & NA & NA \\
& CIFAR-10     & NA & NA & NA & NA & 99.50 \cite{dosovitskiy2021an} & NA & NA \\
& CIFAR-100    & NA & NA & NA & NA & 96.08 \cite{foret2021sharpnessaware} & NA & NA \\
\midrule

\multirow{6}{*}{\cite{salem2019ml}}
& Purchase-100 & NA & $\sim 80.00$ & NA & NA & 90.00~\cite{suri2024parameters} & 10.00 & \xmark \\
& Locations    & NA & $\sim 62.00$ & NA & NA & 72.00 \cite{li2024mist} & 10.00 & \xmark \\
& MNIST        & NA & $\sim 98.00$ & NA & NA & 99.87 \cite{byerly2021no} & 1.87  & \cmark \\
& CIFAR-10     & NA & $\sim 60.00$ & NA & NA & 99.50 \cite{dosovitskiy2021an} & 39.50 & \xmark \\
& CIFAR-100    & NA & $\sim 22.00$ & NA & NA & 96.08 \cite{foret2021sharpnessaware} & 74.08 & \xmark \\
& LFW          & NA & $\sim 68.00$ & NA & NA & 99.83 \cite{deng2019arcface} & 31.83 & \xmark \\
\midrule

\multirow{2}{*}{\cite{sablayrolles2019white}}
& CIFAR-10     & NA & NA & NA & NA & 99.50 \cite{dosovitskiy2021an} & NA & NA \\
& ImageNet-1K  & NA & NA & NA & NA & 94.90 \cite{nikzad2025sata} & NA & NA \\
\midrule

\multirow{3}{*}{\cite{long2020pragmatic}}
& Adult      & 85.00 & 85.00 & 0.00 & \cmark & 88.16~\cite{chakrabarty2018statistical} & 3.16 & \cmark \\
& UCI Cancer & 95.00 & 94.00 & 1.00 & \cmark & 98.86 \cite{breast_cancer} & 4.86 & \cmark \\
& MNIST      & 99.00 & 99.00 & 0.00 & \cmark & 99.87 \cite{byerly2021no} & 0.87 & \cmark \\
\midrule

\multirow{4}{*}{\cite{jayaraman2021revisiting}}
& Purchase-100X & 100.00 & 71.00 & 29.00 & \xmark & 90.00~\cite{suri2024parameters} & 19.00 & \xmark \\
& Texas-100     & 100.00 & 53.00 & 47.00 & \xmark & 57.00 \cite{shokri2017membership} & 4.00  & \cmark \\
& CIFAR-100     & 48.00  & 18.00 & 30.00 & \xmark & 96.08 \cite{foret2021sharpnessaware} & 78.08 & \xmark \\
& RCV1X         & 100.00 & 84.00 & 16.00 & \xmark & 92.85 \cite{johnson2016supervised} & 8.85  & \xmark \\
\midrule

\multirow{4}{*}{\cite{song2021systematic}}
& Purchase-100 & 99.80 & 80.90 & 18.90 & \xmark & 90.00~\cite{suri2024parameters} & 9.10  & \xmark \\
& Texas-100    & 81.00 & 52.30 & 28.70 & \xmark & 57.00 \cite{shokri2017membership} & 4.70  & \cmark \\
& Locations    & 100.00& 60.70 & 39.30 & \xmark & 72.00 \cite{li2024mist} & 11.30 & \xmark \\
& CIFAR-100    & 100.00& 83.00 & 17.00 & \xmark & 96.08 \cite{foret2021sharpnessaware} & 13.08 & \xmark \\
\midrule

\multirow{7}{*}{\cite{liu2022membership}}
& Purchase-100 & NA    & NA    & NA    & NA     & 90.00~\cite{suri2024parameters} & NA    & NA \\
& Locations    & NA    & NA    & NA    & NA     & 72.00 \cite{li2024mist} & NA    & NA \\
& CIFAR-10     & 100.00& 82.60 & 17.40 & \xmark & 99.50 \cite{dosovitskiy2021an} & 16.90 & \xmark \\
& CINIC-10     & 99.90 & 65.80 & 34.10 & \xmark & 95.06 \cite{antonio2025efficient} & 29.25 & \xmark \\
& CIFAR-100    & 99.90 & 47.50 & 52.40 & \xmark & 96.08 \cite{foret2021sharpnessaware} & 48.58 & \xmark \\
& GTSRB        & 100.00& 88.10 & 11.90 & \xmark & 99.71 \cite{arcos2018deep} & 11.61 & \xmark \\
& Newsgroups   & NA    & NA    & NA    & NA     & 89.50 \cite{lin2021bertgcn} & NA    & NA \\
\midrule

\multirow{7}{*}{\cite{watson2022on}}
& Adult         & 90.40 & 84.20 & 6.20  & \cmark & 88.16~\cite{chakrabarty2018statistical} & 3.96  & \cmark \\
& UCI Credit    & 92.10 & 75.10 & 17.00 & \xmark & 83.20~\cite{credit_data} & 8.10  & \xmark \\
& UCI Hepatitis & 99.80 & 87.10 & 12.70 & \xmark & 97.44~\cite{hepatitis_46} & 10.34 & \xmark \\
& MNIST         & 99.00 & 98.90 & 0.10  & \cmark & 99.87 \cite{byerly2021no} & 0.97  & \cmark \\
& CIFAR-10      & 94.60 & 74.00 & 20.60 & \xmark & 99.50 \cite{dosovitskiy2021an} & 25.50 & \xmark \\
& CIFAR-100     & 93.80 & 37.30 & 56.50 & \xmark & 96.08 \cite{foret2021sharpnessaware} & 58.78 & \xmark \\
& ImageNet-1K   & 77.30 & 63.70 & 13.60 & \xmark & 94.90 \cite{nikzad2025sata} & 31.20 & \xmark \\
\midrule

\multirow{4}{*}{\cite{ye2022enhanced}}
& Purchase-100 & 100.00 & 75.50 & 24.50 & \xmark & 90.00~\cite{suri2024parameters} & 14.50 & \xmark \\
& MNIST        & 98.60  & 97.10 & 1.50  & \cmark & 99.87 \cite{byerly2021no} & 2.77  & \cmark \\
& CIFAR-10     & 99.89  & 77.37 & 22.52 & \xmark & 99.50 \cite{dosovitskiy2021an} & 22.13 & \xmark \\
& CIFAR-100    & 97.90  & 20.40 & 77.50 & \xmark & 96.08 \cite{foret2021sharpnessaware} & 75.68 & \xmark \\
\midrule

\multirow{5}{*}{\cite{carlini2022membership}}
& Purchase-100 & NA     & NA    & NA    & NA     & 90.00~\cite{suri2024parameters} & NA    & NA \\
& Texas-100    & NA     & NA    & NA    & NA     & 57.00 \cite{shokri2017membership} & NA    & NA \\
& CIFAR-10     & 100.00 & 90.00 & 10.00 & \cmark & 99.50 \cite{dosovitskiy2021an} & 9.50  & \xmark \\
& CIFAR-100    & 100.00 & 60.00 & 40.00 & \xmark & 96.08 \cite{foret2021sharpnessaware} & 36.08 & \xmark \\
& ImageNet-1K  & 100.00 & 65.00 & 35.00 & \xmark & 94.90 \cite{nikzad2025sata} & 29.90 & \xmark \\
\midrule

\multirow{4}{*}{\cite{bertran2023scalable}}
& CIFAR-10     & NA & 91.0  & NA & NA & 99.50 \cite{dosovitskiy2021an} & 8.50  & \xmark \\
& CINIC-10     & NA & NA    & NA & NA & 95.06 \cite{antonio2025efficient} & NA    & NA \\
& CIFAR-100    & NA & 68.60 & NA & NA & 96.08 \cite{foret2021sharpnessaware} & 27.48 & \xmark \\
& ImageNet-1K  & NA & 67.50 & NA & NA & 94.90 \cite{nikzad2025sata} & 27.40 & \xmark \\
\midrule

\multirow{5}{*}{\cite{zarifzadeh2024low}}
& Purchase-100 & 100.00 & 83.40 & 16.60 & \xmark & 90.00~\cite{suri2024parameters} & 6.60  & \xmark \\
& CIFAR-10     & 99.90  & 92.40 & 7.50  & \cmark & 99.50 \cite{dosovitskiy2021an} & 7.10  & \xmark \\
& CINIC-10     & 99.50  & 77.20 & 22.30 & \xmark & 95.06 \cite{antonio2025efficient} & 17.86 & \xmark \\
& CIFAR-100    & 99.90  & 67.50 & 32.40 & \xmark & 96.08 \cite{foret2021sharpnessaware} & 28.58 & \xmark \\
& ImageNet-1K  & 90.20  & 58.60 & 31.60 & \xmark & 94.90 \cite{nikzad2025sata} & 36.30 & \xmark \\

\bottomrule
\end{tabular}}
\end{table}

\paragraph{\textbf{Condition C3.}}
Table~\ref{tab:metrics} reports the performance metrics of the surveyed attacks and their reliability. 
We derive missing metrics when possible using standard statistical relationships, such as computing FPR from prior, precision, and TPR, or inferring precision from prior, TPR, and FPR. 
These derived values are underlined in the table.
When multiple attack variants are reported (such as in \cite{watson2022on}), we consider the one with the highest precision.
For attacks reporting precision under the balanced prior ($p=50\%$), we recompute the weighted precision with $p=10\%$. 
An attack is considered reliable only if it maintains precision $\geq 95\%$ with this more realistic prior, ensuring minimal false positives. 
When the precision with the balanced prior is 100\% (indicating 0\% FPR), we classify the attack as reliable, since perfect precision is maintained regardless of the prior.

Our analysis reveals that only 16 out of 61 attack-data set pairs (26.2\%) demonstrate reliable membership inference by maintaining high precision with realistic priors. 
Recent approaches like \cite{carlini2022membership} and \cite{zarifzadeh2024low} show particular promise. 
\cite{carlini2022membership} achieve near-perfect precision ($\geq 99.95\%$) with FPR $\leq 0.001\%$ on CIFAR-10, CIFAR-100, and ImageNet-1K, therefore satisfying reliability even with $p = 10\%$.
\cite{zarifzadeh2024low} report 100\% precision and 0\% FPR on all data sets, making it reliable regardless of the prior.
\cite{jayaraman2021revisiting} achieve 100\% precision on CIFAR-100 and RCV1X, while \cite{long2020pragmatic} report 100\% precision on MNIST, but only identify one true member. 
Similarly, \cite{watson2022on} achieve 100\% precision on UCI Hepatitis, CIFAR-10, and CIFAR-100, although the number of true members identified is limited: 1 for UCI Hepatitis, 2.8 ($\pm2.4$) for CIFAR-10, and 3.0 ($\pm1.26$) for CIFAR-100.
This fairly small number of true positives with such high standard deviations adds another layer of uncertainty about the reliability of these attacks.

Earlier attacks such as \cite{shokri2017membership}, \cite{yeom2018privacy}, and \cite{salem2019ml} generally fail to achieve reliable inference, with precision below our threshold due to the high FPR. This is because these attacks are typically optimized for high accuracy.
Some attacks ({\em e.g.}, \cite{sablayrolles2019white}, \cite{ye2022enhanced}) lack critical metrics (precision/FPR), which prevents reliability assessment.
It can also be seen that the attack in \cite{liu2022membership} exhibits substantial drops in precision (for example, 98.65\% to 89.02\% for CIFAR-100) when evaluated with $p=10\%$, which emphasizes the importance of realistic evaluation settings.

In summary, most attacks fail to satisfy condition C3 due to high FPR or precision degradation with realistic priors. Only attacks with 0\% FPR ({\em e.g.}, \cite{zarifzadeh2024low} and \cite{watson2022on} on CIFAR-10/100) or extremely low FPR ($\leq 0.001\%$) paired with $\geq 95.0\%$ precision ({\em e.g.}, \cite{carlini2022membership}) satisfy C3. 

\begin{table}[t!]
\centering
\caption{Evaluation of attack performance and membership reliability (C3)}
\label{tab:metrics}
\resizebox{0.72\textwidth}{!}{
\begin{tabular}{ll|ccccc} 
\toprule
\multicolumn{1}{c}{\multirow{2}{*}{Attack}}& \multicolumn{1}{c}{\multirow{2}{*}{Data set}}& \multicolumn{3}{c}{Attack performance (\%)} & \multirow{2}{*}{Prior $p$ (\%)~} & \multicolumn{1}{c}{\multirow{2}{*}{Reliable?}}  \\
\cmidrule(lr){3-5}
\multicolumn{1}{c}{}                       & \multicolumn{1}{c}{}&        Prec.&            TPR&                FPR&             & \multicolumn{1}{c}{}                \\ 
\hline
\multirow{7}{*}{\cite{shokri2017membership}}    &Adult&                         50.30&          NA&                 NA&             50&                                   \xmark\\
                                                &Purchase-100&                  73.00&          86.00&      \underline{31.81}&      50&                                   \xmark\\
                                                &Locations&                     67.80&          98.00&      \underline{46.54}&      50&                                   \xmark\\
                                                &Texas-100&                     69.00&          86.00&      \underline{38.64}&      50&                                   \xmark\\
                                                &MNIST&                         51.70&          NA&                 NA&             50&                                   \xmark\\
                                                &CIFAR-10&                      71.00&          99.00&      \underline{40.44}&      50&                                   \xmark\\
                                                &CIFAR-100&                     97.00&          99.00&      \underline{3.06}&       50&                                   \xmark (Prec \underline{78.24\%} with $p=10\%$)\\
\hline
\multirow{3}{*}{\cite{yeom2018privacy}}         &MNIST&                         50.50&          99.00&      \underline{97.04}&      50&                                   \xmark\\
                                                &CIFAR-10&                      69.40&          99.00&      \underline{43.65}&      50&                                   \xmark\\ 
                                                &CIFAR-100&                     87.40&          99.00&      \underline{14.27}&      50&                                   \xmark\\    
\hline
\multirow{6}{*}{\cite{salem2019ml}}            
                                                &Purchase-100&                  $\sim55.00$&    $\sim100.0$&  \underline{$\sim81.82$}&    50&                                   \xmark\\
                                                &Locations&                     $\sim80.00$&    $\sim95.00$&  \underline{$\sim23.75$}&    50&                                   \xmark\\
                                                &MNIST&                         $\sim50.00$&    $\sim99.00$&  \underline{$\sim99.00$}&    50&                                   \xmark\\
                                                &CIFAR-10&                      $\sim60.00$&    $\sim15.00$&  \underline{$\sim10.00$}&    50&                                   \xmark\\
                                                &CIFAR-100&                     $\sim90.00$&    $\sim100.00$&  \underline{$\sim11.11$}&   50&                                   \xmark\\
                                                &LFW&                           $\sim70.00$&    $\sim100.0$&   \underline{$\sim42.86$}&   50&                                   \xmark\\                  
\hline
\multirow{2}{*}{\cite{sablayrolles2019white}}   &CIFAR-10&                      NA&            NA&                 NA&             50&                                   NA\\
                                                &ImageNet-1K&                   NA&            NA&                 NA&             50&                                   NA\\
\hline
\multirow{3}{*}{\cite{long2020pragmatic}}       &Adult&                         73.91&          NA&                 NA&             50&                                   \xmark\\ 
                                                &UCI Cancer&                    88.89&          NA&                 NA&             50&                                   \xmark\\  
                                                &MNIST&                         100&            NA&        \underline{0.00}&        50&                         \cmark \\
\hline
\multirow{8}{*}{\cite{jayaraman2021revisiting}} &Purchase-100X&                 98.00&          NA&                 NA&             50&                                   NA\\
                                                &Purchase-100X&                 97.50&          NA&                 NA&             9.0&                          \cmark \\
                                                &Texas-100&                     95.70&          NA&                 NA&             50&                                   NA\\
                                                &Texas-100&                     97.40&          NA&                 NA&             33&                                   NA\\
                                                &CIFAR-100&                     100.0&          NA&                 NA&             50&                           \cmark \\
                                                &CIFAR-100&                     100.0&          NA&                 NA&             33&                           \cmark \\
                                                &RCV1X&                         100.0&          NA&                 NA&             50&                           \cmark \\
                                                &RCV1X&                         93.00&          NA&                 NA&             9.0&                                  \xmark\\
\hline
\multirow{4}{*}{\cite{song2021systematic}}      &Purchase-100&                  63.40&          7.80&         \underline{4.50}&     50&                                   \xmark\\
                                                &Texas-100&                     85.40&          21.20&        \underline{3.62}&     50&                                   \xmark\\
                                                &Locations&                     NA&             NA&                 NA&             50&                                   NA\\
                                                &CIFAR-100&                     NA&             NA&                 NA&             50&                                   NA\\
\hline
\multirow{7}{*}{\cite{liu2022membership}}       &Purchase-100&                  \underline{97.14}&        3.40&                0.10&     50&\xmark (Prec \underline{84.48\%} with $p=10\%$)\\
                                                &Locations&                     \underline{98.00}&        4.90&                0.10&     50&\xmark (Prec \underline{84.00\%} with $p=10\%$)\\
                                                &CIFAR-10&                      \underline{93.75}&        1.50&                0.10&     50&\xmark (Prec \underline{62.50\%} with $p=10\%$)\\
                                                &CINIC-10&                      \underline{98.21}&        5.50&                0.10&     50&\xmark (Prec \underline{85.94\%} with $p=10\%$)\\
                                                &CIFAR-100&                     \underline{98.65}&        7.30&                0.10&     50&\xmark (Prec \underline{89.02\%} with $p=10\%$)\\
                                                &GTSRB&                         \underline{98.65}&        7.30&                0.10&     50&\xmark (Prec \underline{89.02\%} with $p=10\%$)\\
                                                &Newsgroups&                    \underline{96.15}&        2.50&                0.10&     50&\xmark (Prec \underline{73.53\%} with $p=10\%$)\\
\hline

\multirow{7}{*}{\cite{watson2022on}}            &Adult&                         80.00&        NA&                  0.0&              0.50&                                   \xmark\\
                                                &UCI Credit&                    90.00&        NA&                  0.0&              0.50&                                   \xmark\\
                                                &UCI Hepatitis&                 100.0&        NA&                  0.0&              0.50&                \cmark \\
                                                &MNIST&                         66.00&        NA&                  NA&               0.50&                                   \xmark\\
                                                &CIFAR-10&                      100.0&        NA&                  0&               0.50&                 \cmark \\
                                                &CIFAR-100&                     100.0&        NA&                  0&               0.50&                  \cmark \\
                                                &ImageNet-1K&                   NA&           NA&                  NA&              0.50&                                  NA\\
\hline
\multirow{4}{*}{\cite{ye2022enhanced}}          &Purchase-100&                  NA&          NA&                  NA&            0.50&                                   NA\\
                                                &MNIST&                         NA&          NA&                  NA&            0.50&                                   NA\\
                                                &CIFAR-10&                      NA&          NA&                  NA&            0.50&                                   NA\\
                                                &CIFAR-100&                     NA&          NA&                  NA&            0.50&                                   NA\\
\hline
\multirow{5}{*}{\cite{carlini2022membership}}   &Purchase-100&                  \underline{97.62}&   4.10&                0.10&          0.50&  \xmark (Prec \underline{82.00\%} with $p=10\%$)\\
                                                &Texas-100&                     \underline{99.70}&   33.20&               0.10&          0.50&                      \cmark \\
                                                &CIFAR-10&                      \underline{99.95}&  2.20&                0.001&         0.50&                      \cmark \\
                                                &CIFAR-100&                     \underline{99.99}&  11.20&               0.001&         0.50&                      \cmark \\
                                                &ImageNet-1K&                   \underline{99.88}&  0.85&                0.001&         0.50&                      \cmark \\
\hline

\multirow{4}{*}{\cite{bertran2023scalable}}     
                                                &CIFAR-10&                      64.48&        \underline{0.18}&   0.10&         0.50&                                    \xmark\\
                                                &CINIC-10&                      85.46&        \underline{0.59}&   0.10&         0.50&                                    \xmark\\
                                                &CIFAR-100&                     85.41&        \underline{0.59}&   0.10&         0.50&                                    \xmark\\
                                                &ImageNet-1K&                   99.64&        \underline{27.68}&  0.10&         0.50&                         \cmark \\
\hline
\multirow{5}{*}{\cite{zarifzadeh2024low}}       &Purchase-100&                  \underline{100}&      0.41&      0.00&         0.50&                          \cmark \\
                                                &CIFAR-10&                      \underline{100}&      2.13&      0.00&         0.50&                          \cmark \\
                                                &CINIC-10&                      \underline{100}&      1.39&      0.00&         0.50&                          \cmark \\
                                                &CIFAR-100&                     \underline{100}&      3.50&      0.00&         0.50&                          \cmark \\
                                                &ImageNet-1K&                   \underline{100}&      0.06&      0.00&         0.50&                          \cmark \\
\bottomrule
\end{tabular}}
\end{table}

\paragraph{\textbf{Condition C4.}}
Table \ref{tab:feasibility} reports the resource requirements and characterizes the computational cost of the attacks surveyed, categorizing them into overall low, moderate, or high computational cost according to the criteria depicted in Section \ref{sec:condition4}.
We can see that 8 out of 13 attacks (about 62\%) incur high computational costs.
For example, the state-of-the-art attack LiRA \cite{carlini2022membership}, which demands 32 to 256 shadow models, is unaffordable in most realistic scenarios.
RMIA \cite{zarifzadeh2024low} substantially reduces the number of models required (1-4) compared to LiRA, although at the expense of increased query volume.
The approach of \cite{bertran2023scalable} eliminates shadow models, but requires training a deep quantile regression model and substantial hyperparameter tuning, resulting in significant computational overhead that limits its efficiency \cite{bertran2023scalable,zarifzadeh2024low}.
Three attacks fall into the moderate category and require only one additional model and a maximum of 100 queries per sample, while only simple global threshold-based methods \cite{yeom2018privacy,salem2019ml} require no additional models and minimal queries, making their computational cost low.

In summary, although recent advances in MIAs have improved attack reliability, their computational feasibility remains a critical consideration. 
\begin{table}[t!]
\centering
\caption{Evaluation of computational feasibility (C4). Cost factors: M: additional models required, I: inference model complexity, Q: queries required per target sample.}
\label{tab:feasibility}
\resizebox{1.0\textwidth}{!}{
\begin{tabular}{lcccccc} 
\toprule
\multicolumn{1}{l}{Attack} & Resource requirements & \multicolumn{3}{c}{Cost factors} & \multicolumn{1}{c}{Overall} \\
\cmidrule(lr){3-5}
 & & M & I & Q & cost \\
\hline
\cite{shokri2017membership} & 10-100 additional models; $K$ ML attack models; 1 query & High & Moderate & Low & High \\ 
\hline
\cite{yeom2018privacy} & Simple thresholding rule on loss values; 1 query & Low & Low & Low & Low \\  
\hline
\cite{salem2019ml} & Simple thresholding rule; 1K random data points per data set; 1 query & Low & Low & Low & Low \\ 
\hline
\cite{sablayrolles2019white} & 30 additional models; 1 query & High & Low & Low & High \\        
\hline
\cite{long2020pragmatic} & 50 additional models; 1 query & High & Low & Low & High \\                            
\hline
\cite{jayaraman2021revisiting} & 1 additional model; 100 queries & Moderate & Low & Low & Moderate \\               
\hline
\cite{song2021systematic} & 1 additional model; 1 query & Moderate & Low & Low & Moderate \\                          
\hline
\cite{liu2022membership} & 2 additional models; 1 ML attack model; multiple queries & High & Moderate & Moderate & High \\                         
\hline
\cite{watson2022on} & 1 additional model; 1 query & Moderate & Low & Low & Moderate \\ 
\hline
\cite{ye2022enhanced} & 15-999 additional models; 1 query & High & Low & Low & High \\ 
\hline
\cite{carlini2022membership} & 32-256 additional models; under 100 queries & High & Low & Low & High \\ 
\hline
\cite{bertran2023scalable} & Multiple quantile regression models; 1 query & Low & High & Low & High \\ 
\hline
\cite{zarifzadeh2024low} & 1-4 additional models; as per paper, 10\% of training set size is sufficient & High & Low & High & High \\
\bottomrule
\end{tabular}}
\end{table}

\paragraph{\textbf{Overall attack effectiveness.}}
\label{sec:effectiveness}
We assess the overall effectiveness of the surveyed attacks based on their simultaneous fulfillment of conditions C1, C2, C3, and C4, as demanded by our evaluation framework. 
We can see that even with the permissive thresholds we employed, {\em all attacks fail to simultaneously meet all four conditions, and hence no attack is deemed effective.} 
They either target overfitted models (violating C1), use models that significantly underperform w.r.t. the state of the art (violating C2), do not achieve high membership inference precision under realistic priors (violating C3), or have impractical computational costs (violating C4). Any of these violations makes the attack unrealistic in production settings. 
The only attack-data set pairs that meet three of the four conditions are \cite{long2020pragmatic} and \cite{watson2022on} on MNIST.
Specifically, \cite{long2020pragmatic} on MNIST meets C1, C2, and C3 but does not meet C4 because it requires 50 additional models to carry out the attack. On the other hand,
\cite{watson2022on} on MNIST fails to produce reliable inference (C3 not satisfied), but at least meets C1, C2 and C4.
Although the state-of-the-art attacks LiRA~\cite{carlini2022membership} and RMIA~\cite{zarifzadeh2024low} achieve reliable inferences in most data sets (fulfilling C3), they generally do not satisfy the remaining conditions.

\section{Discussion}
\label{sec:discussion}

In the following, we discuss the main limitations and challenges that we observe in current MIAs.

\subsection{Fundamental Limitations of MIAs} 

\rule{15pt}{0pt}\textbf{The membership paradox.} Our analysis reveals a fundamental paradox in membership inference: when training data are non-exhaustive (as most real-world training sets are), there exists plausible deniability and, therefore, intrinsic protection against any 
MIA-enabled attribute disclosure; conversely, when training data are exhaustive and attribute disclosure might be more certain, membership is already known and MIAs become irrelevant. 
This undermines the privacy threat posed by MIAs in a way that is agnostic of the attack design.
Although non-unequivocal inferences with moderate precision under skewed priors may provide leads for further investigation, they still substantially weaken the certainty of individual membership claims and provide plausible deniability to data subjects.

\textbf{Overfitting vs. practical deployment.} We must distinguish between intentional overfitting ---where models are deliberately overfitted to facilitate MIAs, as it was the case in several of the reviewed works---, and unavoidable overfitting ---which arises naturally due to data set complexity---. 
Our framework focuses on privacy risks exploitable by realistic attackers, rather than artifacts of data or training processes. 
State-of-the-art MIAs often exaggerate their effectiveness by training target models on small subsets of available data and reserving large portions (\emph{e.g.}, 50\%) for shadow models. 
This practice induces overfitting and deviates significantly from real-world ML practices, where models are trained on large, diverse data sets with the ultimate goal of maximizing generalization.

\textbf{Data diversity and size mitigate risks.} Increasing the diversity and size of training data sets generally reduces the precision of MIAs~\cite{shokri2017membership}. 
Modern ML systems, particularly in high-stakes domains, typically utilize large and diverse training sets, which inherently provides protection against membership inference.

\subsection{Methodological Issues in Current MIA Research}

\rule{10pt}{0pt}\textbf{Reference data dependence.} A critical limitation of state-of-the-art MIAs is their reliance on reference data sets that closely mirror the target model's training distribution. 
Attacks like LiRA~\cite{carlini2022membership} require shadow models trained on data similar to the target private data set. 
As demonstrated by~\cite{fu2024membership}, this dependence leads to high false positive rates when reference data differ, which is a common scenario when private or proprietary data sets are used to fine-tune modern ML models such as large language models (LLMs).

\textbf{Unrealistic membership priors.} The reliability of MIAs is further undermined by realistic membership priors, where non-members vastly outnumber members. 
As shown by~\cite{watson2022on}, increasing the number of non-member samples improves attack accuracy by favoring non-membership inference, but this comes at the cost of degraded precision-recall trade-offs. 
In real-world scenarios with low membership priors (10\% or less), achieving high precision often results in low recall, limiting the attack's ability to reliably identify members. 
This highlights the need to evaluate MIAs under realistic priors rather than the unrealistic balanced settings (50\% of members) commonly used in the literature.

\textbf{Inconsistent positive inference.} For MIAs to be reliable, positive membership inferences must remain consistent across different random initializations and small variations in training data. 
However,~\cite{ye2022enhanced} observe that the same target sample does not consistently produce a positive membership inference across different target models trained with varying initializations or slight changes in training data. 
Even when MIAs report high precision, their practical impact is limited by inconsistent results and low coverage. 
High standard deviations in multiple runs mean that \emph{different} records are flagged as vulnerable depending on the initialization of the model or hyperparameters, thus reducing confidence in any single inference.

\textbf{Reproducibility concerns.} The previous issues are compounded by reproducibility challenges that further erode confidence in the reported performance of MIAs. 
For example, when reproducing the offline LiRA attack using the authors' own code, \cite{zarifzadeh2024low} observed significantly lower attack accuracy than reported in \cite{carlini2022membership}. 
This inconsistency may be attributed to different training initializations or data splitting approaches, highlighting the instability of current MIA techniques.

\subsection{Implications for Privacy-preserving ML}

\rule{15pt}{0pt}\textbf{Unjustified utility loss.} Given that MIAs pose a weak privacy threat (or none at all) under realistic conditions, the utility cost of defenses to protect
against MIAs, such as DP or other noise addition techniques, appears disproportionate and unwarranted. 
In critical domains such as healthcare, even a 1\% drop in performance can have serious consequences, particularly when diagnosing rare diseases. 
Therefore, practitioners should carefully weigh the actual privacy risks (if any) against utility losses before implementing systematic protection measures.

\textbf{Targeted disclosure protection.} Given that attribute disclosure risks come from training data features (condition C0), instead of applying general mechanisms to counter MIAs, a better balance between privacy and utility would be achieved by employing safeguards especially designed to break disclosure-enabling data features. Specifically, data sampling counteracts the exhaustiveness of the training data set \cite{sdc-book}, whereas formalisms such as $l$-diversity \cite{ldiversity} break confidential attribute uniqueness by introducing a minimum variance in confidential attributes. 
Other measures may involve replacing outlying or otherwise vulnerable attribute values with synthetic imputed values (partial synthesis). 
Compared to DP, which systematically \emph{distorts all data or model parameters with some probability}, the aforementioned methods \emph{target specific data points or data features with certainty}. 

\textbf{Regulatory implications.} Our findings suggest that current regulatory frameworks may overestimate the privacy risks associated with ML models. 
Unless MIAs are significantly improved or
new privacy-meaningful attacks are devised, the alleged privacy risks of predictive ML seem mild enough to warrant reconsideration of stringent privacy requirements that might hinder innovation and utility.

\subsection{Towards Improved Evaluation Standards}

\rule{10pt}{0pt}\textbf{Standardized reporting practices.} The lack of consistent and comprehensive reporting in MIA research hampers meaningful comparisons between studies. 
Critical metrics, such as training and test loss curves during training, accuracy/loss gaps after training, the number of true positives from member samples, and the stability of the results, are often omitted. 
We advocate for future studies to consistently report these metrics, along with runtime costs and standard deviations across multiple runs with realistic priors. Only this will enable a realistic evaluation of whether MIAs are effective or simply succeed under contrived conditions.

\textbf{Alternative privacy metrics.} Given the limitations of MIAs, rather than focusing exclusively on them as a proxy for privacy risk assessment, researchers should explore alternative metrics to assess meaningful privacy risks in ML, such as re-identifiability and actual attribute disclosure. 
These might include measures of information leakage that do not rely on binary membership inference, evaluations of model vulnerability to reconstruction attacks, or assessments of how model outputs correlate with sensitive attributes in the training data.
This balanced assessment would help avoid unnecessary protection measures while ensuring appropriate safeguards where genuinely needed.

\section{Conclusions, Recommendations, and Future Research}
\label{sec:conclusion}

MIAs are the main method to assess privacy risks in ML and a building block for other privacy attacks such as data extraction or reconstruction. In this work, we have shown that MIAs are unlikely to disclose sensitive information. On the one hand, the data used to train the target model must satisfy very stringent conditions such as exhaustivity and uniqueness of confidential attribute values; otherwise, any attribute disclosure is plausibly deniable. On the other hand, state-of-the-art MIAs fail to offer reliable performance on non-overfitted, competitive target models at a reasonable computational cost.  

Thus, our analysis demonstrates that MIAs generally do not constitute significant privacy threats under realistic conditions. 
Hence, relying on MIAs as primary privacy risk metrics can lead to overestimation of privacy vulnerabilities and unjustified utility loss caused by unnecessary protection techniques. This finding has important implications for both research directions and practical privacy-preserving ML deployment.

A broader ramification is that, unless MIAs are improved or more effective privacy attacks are devised, the alleged privacy risks of predictive ML seem mild enough to consider a relaxation of regulatory frameworks.

Recommendations to practitioners that release trained models include:
\begin{itemize}
\item Check whether the data that will be used to train the model to be released satisfy the necessary conditions listed in Section~\ref{sec:disclosure_risk} for the unequivocal disclosure of sensitive information. 
\item If they do not, then refrain from using privacy protection techniques that may result in unwarranted utility loss.
\item If they do, resort to privacy-preserving mechanisms that target problematic data features (such as sampling or partial synthesis), and make sure that the trained model is not overfitted. To fix overfitting, standard anti-overfitting techniques should offer a better privacy-utility trade-off than DP~\cite{blanco2022critical}.  
\end{itemize}

As future work, we plan to analyze the potential of the privacy attacks proposed for generative models, including large language models and diffusion models, which have been shown to memorize and leak data from the training set \cite{carlini2021extracting}. For these models, it is less obvious how to formulate and check conditions on the training data, which do not consist of a single specific data set.

\begin{credits}
\subsubsection{\ackname} 
Thanks go to Alberto Blanco-Justicia for discussions on earlier versions of this paper. Partial support for this work has been received from 
the Government of Catalonia (ICREA Acad\`emia Prizes to J. Domingo-Ferrer and to D. S\'anchez), MCIN/AEI under grant PID2024-157271NB-I00 ``CLEARING-IT'', and the European Commission under project
HORIZON-101292277 ``SoBigData IP''.\\ \\
\end{credits}

\appendix

This appendix provides supplementary material supporting the main text. 
It includes detailed descriptions of the surveyed attacks and additional background information.

\section{Surveyed Attacks}
\label{supp:surveyed-attacks}

This section provides the detailed survey referenced in the main manuscript. In particular, 
Table~\ref{tab:summary}
summarizes the representative black-box membership inference attacks (MIAs) reviewed in the paper, including
their core ideas, publication venues, and citation counts.

\begin{table}[!ht]
\centering
\caption{Summary of the surveyed attacks. Citation counts according to Google Scholar, January~8,~2026.}
\label{tab:summary}
\resizebox{\textwidth}{!}{
\begin{tabular}{lccc} 
\toprule
\multicolumn{1}{c}{Attack}                              & Approach & Venue                                   & \# citations  \\ 
\hline
\cite{shokri2017membership}    &    \multicolumn{1}{l}{ML membership classifier on predictions from shadow models}  & IEEE SP~2017                            & 7,083          \\ 
\hline
\cite{yeom2018privacy}         &   \multicolumn{1}{l}{Loss global thresholding}   & IEEE CSF~2018                                           & 1,751          \\ 
\hline
\cite{salem2019ml}         &   \multicolumn{1}{l}{Confidence and entropy global thresholding}   & NDSS ~2019                                & 1,329          \\ 
\hline
\cite{sablayrolles2019white}   &   \multicolumn{1}{l}{Per-sample loss calibration and thresholding}   & ICML 2019                           & 515           \\ 
\hline
\cite{long2020pragmatic}       &   \multicolumn{1}{l}{Hypothesis testing on loss values of selected vulnerable records}   & Euro SP 2020    & 143            \\ 
\hline
\cite{jayaraman2021revisiting} &  \multicolumn{1}{l}{Perturbed input loss thresholding}   & PoPETs~2021                 & 194           \\ 
\hline
\cite{song2021systematic}      &    \multicolumn{1}{l}{Class-specific modified entropy thresholding}   & USENIX Security 2021        & 574           \\ 
\hline
\cite{liu2022membership}       &  \multicolumn{1}{l}{ML membership classifier on the sample's loss of trajectory from distilled models}    & ACM~CCS 2022   & 165            \\ 
\hline
\cite{watson2022on}            &   \multicolumn{1}{l}{Per-sample loss calibration and thresholding}   & ICLR 2022                      & 203           \\ 
\hline
\cite{ye2022enhanced}          &   \multicolumn{1}{l}{Hypothesis testing on loss values from reference/distilled models}   & ACM CCS 2022       & 429           \\ 
\hline
\cite{carlini2022membership}   &   \multicolumn{1}{l}{Hypothesis testing based on likelihood ratio of scores from shadow models}   & IEEE SP 2022             & 1180           \\ 
\hline
\cite{bertran2023scalable}     &   \multicolumn{1}{l}{Quantile regression on confidence scores}   & NeurIPS 2024                                 & 81            \\ 
\hline
\cite{zarifzadeh2024low}       &   \multicolumn{1}{l}{Hypothesis testing based on likelihood ratio of scores from shadow models}   & ICML 2024                 & 91             \\
\bottomrule
\end{tabular}}
\end{table}

The first notable MIA against DNN models was presented in~\cite{shokri2017membership}, and involves training multiple ``shadow'' models on data sampled from a distribution similar to that of the target model to replicate its behavior. 
The prediction outputs of these shadow models on both their training and non-training data are labeled according to membership status. 
These labeled outputs are then used to train attack models that learn to distinguish members from non-members based on patterns in the model outputs.

\cite{salem2019ml} relax some assumptions made by \cite{shokri2017membership} and demonstrate that MIAs could be conducted using one of the following threat models: 
1) only a single shadow model and prior knowledge of training data distribution (Adversary 1); 2) a single shadow model and no prior knowledge of the target model's architecture or training data distribution (Adversary 2); or 3) no need to train shadow models at all (Adversary 3). 
For Adversary 3, the authors introduce label- and shadow model-free attacks based on the prediction entropy of the target model \( f_{\theta}(x) \) or the maximum confidence score assigned by the target model for a given input \( x \). In these cases, it is suggested that the membership status decision be made by using a global threshold $\tau$, such as percentiles ({\em e.g.}, the top 10\%), based on model output statistics derived from random or synthetic data points representing non-members.

\cite{yeom2018privacy} explore the privacy risks related to overfitting and theoretically show that higher generalization errors make models more vulnerable to MIAs, as members often exhibit higher prediction confidence or lower loss values compared to non-members.
They propose several simple and low-cost MIAs based on the generalization gap of the target model. 
Among them, a method computes the loss of the target model \( f_{\theta} \) on \( (x,y) \), and if the loss is below the expected training loss, the point is inferred to be a member.

Score-based attacks proposed by~\cite{yeom2018privacy,salem2019ml} exploit the training objective of minimizing prediction loss in training data, which often results in training samples that achieve near-maximal confidence for their true labels, while test samples exhibit lower confidence. 
In such cases, a global threshold \( \tau \) can be applied to infer membership: samples with confidence exceeding \( \tau \) (or loss below \( \tau \)) are classified as members.

\cite{song2021systematic} demonstrate that score-based approaches can be improved using class-specific thresholds $\tau_y$ (for a class label $y$) instead of a single global threshold for loss values. 
The intuition is that an unbalanced data set can cause the target model to exhibit varying confidence levels across different class labels. 
The class-dependent thresholds $\tau_y$ are learned by training a shadow model to mimic the behavior of the target model, collecting the shadow model's metric values ({\em e.g.}, prediction confidence) on both shadow training and test data, and selecting $\tau_y$ to maximize the accuracy of distinguishing between members and non-members of the shadow model for class $y$.
Additionally, they propose using a modified prediction entropy of the sample as a metric, which incorporates information about the ground-truth label.

\cite{jayaraman2021revisiting} leverage the observation that training samples are typically near a local minimum of the loss function of the model. 
Their attack perturbs an input \(x\) with fresh Gaussian noise, queries the model on perturbed inputs, and counts how often the perturbed inputs result in a higher loss than that of \((x, y)\), 
where $y$ is the class label. 
If this count exceeds a specified threshold $\tau$, \((x, y)\) is classified as a member. 
To define the threshold, a shadow model is trained on data sampled from the target model data distribution, loss values are collected on shadow training/test data, and a global threshold $\tau$ is selected that maximizes TPR while constraining FPR to a desired level ({\em e.g.}, $\alpha=1\%$).
Unlike prior attacks that assume an unrealistic balanced membership prior (\(p = 0.5\)), they evaluate the attack precision under a skewed prior \(p \ll 0.5\), 
because in practical scenarios members are often a small subset of a broader population. 

Other works focus on addressing a significant challenge shared by the above-mentioned MIAs: the high false-positive rate, which undermines the reliability of these attacks because they reduce confidence in the attack's predictions. 

\cite{long2020pragmatic} aim to improve the precision of the attack by selectively targeting ``vulnerable'' samples that have a unique influence on the target model.
To identify these samples, the authors compute the cosine similarity between the feature representations of the data points and select the top 10\% with the greatest distances from their nearest neighbors.
The intuition here is that samples with fewer neighbors are more likely to impose a unique influence on the model.
After that, they train multiple reference models on data sets that exclude these vulnerable samples to estimate the loss distribution for non-members.
Membership inference is then performed using a statistical hypothesis test.
The attacker queries the target model with a record and computes a p-value under the null hypothesis that the record is a non-member, based on the estimated non-member loss distribution.
If the p-value falls below a predefined threshold ({\em e.g.}, 0.01), the record is classified as a member.

In practical scenarios, well-generalized target models typically exhibit similar behavior on member and non-member samples, making it challenging to differentiate between them. 
To address this issue, \cite{liu2022membership} exploit the differences in the training loss trajectories of the member and non-member samples. 
First, they use knowledge distillation to estimate the training trajectory of the target model. 
Then, they train a binary ML classifier using the concatenated loss values from the student model across training epochs, along with the loss from the target model, to capture membership patterns and distinguish between members and non-members.

\cite{sablayrolles2019white} demonstrate that the optimal MIA relies solely on a model's loss function, with white-box access providing no additional information beyond the loss itself.  
The attack assigns a sample-specific score
\(\mathcal{A}'(x,y) = -\mathcal{L}(f_\theta(x), y) + \tau_{x,y},\)
where the calibration term \( \tau_{x,y} \) accounts for the inherent prediction difficulty of \((x,y)\) when excluded from training.  
A low \( \tau_{x,y} \) indicates that \((x,y)\) is naturally easy to predict, which means that a low sample loss from the target model does not necessarily imply positive membership.  
To approximate \(\tau_{x,y}\), the authors train multiple shadow models, half of them including \((x,y)\) in their training set (IN models) and half excluding it (OUT models).  
They then determine the threshold that best separates the loss distributions of the IN and OUT models for each sample.

To reduce computational cost, \cite{watson2022on} approximate the difficulty calibration by setting the term \( \tau_{x, y} \) to the average score (based on metrics such as loss or confidence) calculated from a set of OUT shadow models that do not include \( (x, y) \) in their training data. 
The intuition is that some non-member samples may still exhibit high membership scores due to their being easy to predict, which results in high false positives. 
Hence, if a sample is easy to predict both for the target model and for the shadow models, its membership score should be reduced accordingly. 
\cite{watson2022on} demonstrate that this approach can improve the inference performance of several existing attacks, including those based on loss, gradient norm, and confidence scores.

\cite{ye2022enhanced} designed sample-dependent (Attack R), and model-and-sample-dependent (Attack D) MIAs by training $N$ shadow/reference/distilled models. In both attacks, the target sample $(x, y)$ is excluded from their training data. 
Inference is performed by conducting an one-sided hypothesis test on the losses computed on $(x, y)$ by the $N$ models.
To target a specific FPR $\alpha$, their attack sets the decision threshold (depending on the attack, on the target model and/or sample) so that a fraction $\alpha$ of the measured losses lies below the threshold.

\cite{carlini2022membership} introduced the Likelihood Ratio Attack (LiRA), which improves over \cite{sablayrolles2019white,ye2022enhanced} by modeling the confidence scores on a given sample \((x, y)\) from multiple IN and OUT shadow models as Gaussian distributions using logit scaling trick. 
It performs a parametric likelihood-ratio test between the two distributions to infer membership, with thresholds set to achieve a desired low FPR. 
Two variants of the attack are presented: online and offline. 
The online version trains IN and OUT shadow models for each target sample, which allows for precise modeling but incurs significant computational cost. 
To mitigate this limitation, the offline version pretrains only the OUT shadow models and measures the probability of observing a score as high as the target model in the OUT scores distribution.

Despite the effectiveness of \cite{sablayrolles2019white,ye2022enhanced,watson2022on,carlini2022membership} in achieving high TPR at low FPR, their computational costs render them unscalable for practical privacy auditing~\cite{zarifzadeh2024low}.

Recent methods by \cite{bertran2023scalable} and \cite{zarifzadeh2024low} aim to address the scalability limitations of LiRA and related attacks. 
\cite{zarifzadeh2024low} demonstrated that, under practical computational budgets (e.g., two shadow models), LiRA's performance degrades to near-random guessing, while Attack-R~\cite{ye2022enhanced} shows low true-positive rates (TPR) at low FPR.

\cite{bertran2023scalable} proposed an attack that requires a single regression model and no knowledge of the architecture of the target model. 
Their approach involves querying the target model using a data set not included in its training to obtain confidence scores. 
Then a quantile regression model is trained on these confidence scores to predict the \(1-\alpha\) quantile of the confidence score distribution, allowing input-specific thresholds for membership inference.

\cite{zarifzadeh2024low} introduced RMIA, which constructs a membership score by comparing the likelihood of the model's confidence scores when \((x, y)\) is in the training set versus when a random data point \(z\) is used. 
The authors show that RMIA can achieve comparable membership detection with fewer reference models (between 1 and 4 models) than LiRA.

\section{Background}
\label{sec:background}
\subsection{Machine Learning}
\label{sec:ml_dnn}

A classification machine learning model \( f_\theta : \mathcal{X} \rightarrow [0, 1]^C \) maps an input \( x \in \mathcal{X} \) to a probability distribution over \( C \) classes, where \( f_\theta(x)_y \) denotes the predicted probability for class \( y \)~\cite{goodfellow2016deep}.  
Given a training set \( D_{\text{train}} \) drawn from an underlying distribution \( \mathbb{D} \), the goal is to train \( f_\theta \) so that it generalizes well to an unseen test set \( D_{\text{test}} \sim \mathbb{D} \).  
Generalization is usually measured by the difference in performance ({\em e.g.}, accuracy or loss) between the training and test sets.  

Our survey focuses on MIAs targeting classification deep neural networks (DNNs), as they are among the most commonly used in the literature.  

A DNN model can be represented as $f(x) = \sigma(z(x))$, where \( z : \mathcal{X} \rightarrow \mathbb{R}^C \) returns unnormalized logits, followed by the softmax function \( \sigma( \cdot) \), which converts logits into a probability vector of length \( K \).  
DNNs are commonly trained using stochastic gradient descent (SGD)~\cite{lecun1998gradient} to minimize a chosen loss function \( \mathcal{L} \).  
For classification tasks, the cross-entropy loss is widely used, that is,
\[ \mathcal{L} = -\sum_{i=1}^C y_i \log(p_i),\]
where $C$ is the number of classes, $y_i$ is the true
probability for class $i$ (usually 1 for the true
class and 0 for all others), and $p_i$ is the predicted 
probability for class $i$.

\subsection{Differential Privacy}
\label{sec:dp}
Differential privacy (DP)~\cite{dwork2006differential} is a privacy framework designed to protect individual contributions within a data set by adding calibrated noise to the outputs. 
Formally, a mechanism \( \mathcal{M} \) satisfies \((\epsilon, \delta)\)-DP if, for any two neighboring data sets \( D \) and \( D' \) differing by a single entry, and for any subset \( S \subseteq \text{Range}(\mathcal{M}) \), it holds that

\begin{equation}
\Pr[\mathcal{M}(D) \in S] \leq e^{\epsilon} \Pr[\mathcal{M}(D') \in S] + \delta,
\end{equation}

where the privacy budget \( \epsilon \) controls the level of privacy, with smaller values indicating stronger privacy guarantees, and \( \delta \) is the probability that the privacy guarantee may fail.

Originally proposed to protect released statistical data, DP has become the standard method to enhance privacy in a variety of applications, including data releases~\cite{models} and privacy-preserving ML~\cite{liu2021machine}. 
In particular, DP is recognized as a rigorous defense against MIAs 
because, if meaningfully applied, it limits the influence of \emph{any} single training data point on the resulting model, thus mitigating the risk that its membership can be inferred~\cite{yeom2018privacy}. 

\subsection{Privacy-preserving Model Training}
\label{sec:privacy_preserving_training}

To enhance the privacy of DNNs, two main training-time approaches have been employed: DP and anti-overfitting techniques. 
Incorporating DP in DNN training often involves modifying the SGD algorithm by clipping the gradients and adding calibrated noise to the clipped gradients~\cite{abadi2016deep}. 
Many works have enforced DP on DNN models to mitigate MIAs~\cite{yeom2018privacy,ying2020privacy,choquette2021label,humphries2023investigating,jayaraman2019evaluating,jayaraman2021revisiting,li2021membership,rahman2018membership}.
Although DP theoretically offers formal privacy guarantees, its practical implementation cannot afford meaningful values (\emph{i.e.}, small enough) of its privacy budget
\(\epsilon\) without significantly degrading the utility of trained models~\cite{jayaraman2019evaluating,blanco2022critical}.

Anti-overfitting techniques, on the other hand, provide empirical privacy protection (without formal guarantees) while better retaining (or even improving) model utility~\cite{blanco2022critical,carlini2022membership}. 
These defenses include methods such as weight regularization~\cite{krogh1991simple}, weight dropout~\cite{srivastava2014dropout}, data augmentation~\cite{van2001art,zhong2020random}, learning rate tuning~\cite{jacobs1988increased}, and early stopping~\cite{prechelt1998early}.

\subsection{Membership Inference Attacks}
\label{sec:MIA}
Membership inference attacks (MIAs) aim to determine whether a given sample $(x, y)$ was part of the training data of a target model $f_{\theta}$~\cite{shokri2017membership}, where $x$ denotes the input feature and $y$ the corresponding class label.
Black-box MIAs only require the ability to query the model and observe its outputs, whereas white-box attacks need full access to the internal parameters of the model~\cite{nasr2019comprehensive,hu2022membership}.
Due to this stringent requirement, approaches based on white-box MIAs are often unfeasible in real-world scenarios where attackers usually only have black-box access to the model. 
On the other hand, as shown in~\cite{sablayrolles2019white}, the optimal attack strategy is primarily based on the loss function of the model, making black-box attacks nearly as effective as their white-box counterparts. 
Due to these reasons, in this work we focus on black-box MIAs, which can be formalized as a security game between a challenger $\mathcal{C}$ and an attacker $\mathcal{A}$ following priorworks~\cite{yeom2018privacy,jayaraman2021revisiting,carlini2022membership}.

\vspace{-0.25cm}
\begin{definition}[Membership Inference Security Game]
\leavevmode\newline
\vspace{-0.5cm}
\begin{enumerate}
    \item $\mathcal{C}$ samples a dataset $D_{\text{train}} \leftarrow \mathbb{D}$ and trains a model $f_\theta$ on $D_{\text{train}}$.
    \item $\mathcal{C}$ randomly samples $b \in \{0,1\}$, where $b=1$ with probability $p$ (most MIAs assume $p=0.5$).
    \item If $b=1$, $\mathcal{C}$ samples $(x,y)$ from $D_{\text{train}}$; otherwise, $(x,y)$ is sampled from $\mathbb{D} \setminus D_{\text{train}}$.
    \item $\mathcal{C}$ gives $\mathcal{A}$ access to $(x,y)$ and query access to $f_\theta$, as well as potential access to the data distribution $\mathbb{D}$.
    \item The attacker outputs $\hat{b} = \mathcal{A}(x,y)$.
    \item The game outputs 1 if $\hat{b} = b$, and 0 otherwise.
\end{enumerate}
\end{definition}

Instead of directly outputting a binary decision, $\mathcal{A}$ often computes a score $\mathcal{A}'(x,y)$ based on metrics such as the model's loss or prediction confidence, which is thresholded at $\tau$ to produce the final membership prediction:

\begin{equation}
\mathcal{A}(x,y) = \mathbf{1}[\mathcal{A}'(x,y) > \tau].
\end{equation}

The attack outcomes are true positive (TP) when $b=1$ and $\hat{b}=1$, false positive (FP) when $b=0$ and $\hat{b}=1$, true negative (TN) when $b=0$ and $\hat{b}=0$, and false negative (FN) when $b=1$ and $\hat{b}=0$.

\paragraph{\textbf{Evaluation Metrics.}} Evaluating MIAs’ performance involves a variety of metrics commonly used in binary classification.
The true positive rate \textit{TPR}, also called \textit{recall}, is computed 
as $TP/(TP+FN)$, and is the fraction of real members correctly identified. The false positive rate \textit{FPR} 
is computed as $FP/(FP+TN)$, and is the fraction of non-members incorrectly classified as members (false alarms). 

\textit{Accuracy (Acc)}, computed as $(TP+TN)/(TP+TN+FP+FN)$, measures overall correct predictions, while the area under the ROC curve (\textit{AUC}) measures the ranking quality of members 
over non-members across various thresholds, with $AUC=1$ being
perfect classification and $AUC=0$ being random classification.

Although \textit{Acc} and \textit{AUC} provide measures of average-case performance, they can overlook the reliability of positive predictions. 
In particular, \textit{AUC} aggregates performance for all
{\em FPRs}, including regions ---such as \textit{FPR} $>$ 10\%--- that are practically irrelevant, since the \textit{TPR} in these regions does little to capture the efficacy of real-world attacks~\cite{rezaei2021difficulty,carlini2022membership,watson2022on}. 
Similarly, optimizing for accuracy can inadvertently inflate \textit{FPR}, thereby compromising the reliable detection of membership~\cite{watson2022on}.

\textit{Precision (Prec)}, defined as \(TP/(TP + FP)\), indicates the reliability of positive predictions, though a high precision value may co-occur with an extremely low \textit{TPR} if many members are missed. 
The \textit{F1 score} balances precision and recall, yet it can mask the individual trade-offs between the two. 
\textit{Membership advantage (MA)} is the difference between \textit{TPR} and \textit{FPR} and quantifies the improvement over random guessing. 
However, high \textit{MA} could occur with non-negligible \textit{FPR}, and it also can overestimate privacy risks under imbalanced priors~\cite{jayaraman2021revisiting}.

Given their widespread use in privacy auditing, MIAs must be evaluated based on their ability to reliably detect membership. Recent state-of-the-art works~\cite{ye2022enhanced,carlini2022membership,bertran2023scalable,zarifzadeh2024low} focus on measuring \textit{TPR} at extremely low \textit{FPR} (FPR $\leq 1\%$) to ensure the reliability of positive detections. 
However, this approach overlooks the typically low prior probability of membership, since the training set often represents a small subset of the overall population. 
As noted in~\cite{jayaraman2021revisiting}, traditional precision does not account for this realistic imbalance. 
For example, during an epidemic, the training set may consist of hospitalized patients with symptoms, while the non-member population comprises the broader city. 
To address this issue, they proposed a weighted precision metric that incorporates the prior membership probability, \(p\), as follows:
\begin{equation}
\label{eq:prec2}
\text{Prec} = \frac{p \times \text{TPR}}{p \times \text{TPR} + (1-p) \times \text{FPR}},
\end{equation}
where \(p \ll 50\%\) in real-world scenarios.

\bibliographystyle{splncs04}
\bibliography{mybibliography}

\end{document}